\documentclass[11pt,letterpaper]{article}

\usepackage[T1]{fontenc}
\usepackage{graphicx}
\usepackage[dvips]{epsfig}
\usepackage{amssymb,amsbsy,amsmath,amsfonts,amssymb,amscd}
\usepackage{multicol}
\usepackage{anysize}
\usepackage{setspace}

\usepackage{setspace}

\usepackage[round,colon]{natbib}
\newlength{\textlarg}

\newcommand{\Vs}{V_{\text{\sc f}}}
\newcommand{\Vl}{V_{\text{\sc c}}}
\newcommand{\Vf}{V_{{\ell}}}
\newcommand{\ds}{d_{\text{\sc f}}}
\newcommand{\dl}{d_{\text{\sc c}}}
\newcommand{\phis}{\phi_{m}^{\text{{\sc f}-dom}}}
\newcommand{\phil}{\phi_{m}^{\text{{\sc c}-dom}}}
\newcommand{\betas}{\phi_m^{\text{\sc f}}}
\newcommand{\betal}{\phi_m^{\text{\sc c}}}
\newcommand{\als}{a_{\text{\sc cf}}}
\newcommand{\bsl}{b_{\text{\sc fc}}}

\author{Thai-Son Vu, Guillaume Ovarlez and
Xavier Chateau~\footnote{Corresponding author;
Electronic mail: xavier.chateau@lcpc.fr} \\
Universit\'e Paris-Est, Laboratoire Navier, CNRS, \\
2 all\'ee Kepler,
77420 Champs sur Marne, France.}
\title{Macroscopic behavior of bidisperse suspensions of noncolloidal
 particles in yield stress fluids}
\date{}

\begin{document}
\maketitle

\renewcommand{\abstractname}{Synopsis}
\begin{abstract}
We study both experimentally and theoretically the
rheological behavior of isotropic bidisperse suspensions of
noncolloidal particles in yield stress fluids. We focus on
materials in which noncolloidal particles interact with the
suspending fluid only through hydrodynamical interactions. We
observe that both the elastic modulus and yield stress of
bidisperse suspensions are lower than those of monodisperse
suspensions of same solid volume fraction. Moreover, we show that
the dimensionless yield stress of such suspensions is linked to
their dimensionless elastic modulus and to their solid volume
fraction through the simple equation
of~\cite{Chateau-Ovarlez-Luu-2008}. We also show that the effect
of the particle size heterogeneity can be described by means of a
packing model developed to estimate random loose packing of
assemblies of dry particles. All these observations finally allow
us to propose simple closed form estimates for both the elastic
modulus and the yield stress of bidisperse suspensions: while the
elastic modulus is a function of the reduced volume fraction
$\phi/\phi_m$ only, where $\phi_m$ is the estimated random loose
packing, the yield stress is a function of both the volume
fraction $\phi$ and the reduced volume fraction.
\end{abstract}

\section{Introduction}
Many industrial processes such as concrete casting, drilling muds,
food-stuff transport ... and natural phenomena, such as slurries,
lava flows ... involve suspensions of polydisperse particles
suspended in a non-Newtonian fluid. Knowing and predicting the
rheological properties of such suspensions is thus a major issue
of both industrial materials mix design and science of deformation
and flow of materials. For instance, it is well known that the
grading of aggregate is one of the main factor influencing the
hardened concrete strength; on the other hand, it has a large
impact on the workability of concrete in its fresh
state~[\cite{Neville-1981}]. Consequently, a lot of work has been
devoted to predict the influence of the particle size distribution
on the concrete overall properties, both in the hardened and fresh
state, and then to elaborate rational mix design
methods~[\cite{de-Larrard-1999}]. As the rheological properties of
a suspension (viscosity, yield stress, ...) are increasing
functions of the solid particle volume fraction, it is necessary
to lower this effect in order to design concrete mixtures
containing the maximum amount of solid particles possible that can
be transported, placed and finished easily. For this purpose, it
is better to use aggregate with a distributed grading rather than
particles of similar sizes.

The behavior of polydisperse, in particular bidisperse, Newtonian
suspensions has received large interest from rheologists and
several works, either theoretical or experimental, have been
devoted to this subject in the last few decades. Different
theoretical relationships aiming at predicting the viscosity of
polydisperse suspensions as a function of the solid volume
fraction $\phi$ and of the particle size distribution have been
proposed in the
literature~[\cite{Farris-1968,Chong-Christiansen-Baer-1971,
Storms-Ramarao-Weiland-1990,Gondret-Petit-1997}]. In the three
last works, the influence of the particle size distribution is
taken into account through the concept of {\em maximum packing
fraction} even if this concept is not always rigorously defined.
It seems quite natural to define the  maximum packing fraction of
a suspension as the volume fraction for which the rheological
properties tend to diverge. { Nevertheless, the determination
of the value of the solid volume fraction at which the rheological
properties diverge, and its link to packing models, pose problem.}
In the case of monodisperse spheres, it seems to be equal to
$0.57$ for isotropic dispersions of
particles~[\cite{Mahaut-Chateau-Coussot-Ovarlez-2008}], to $0.605$
for anisotropic dispersions~[\cite{Ovarlez-Bertrand-Rodts-2006}]
while the random close packing of the dry particles is equal to
$0.64$.

\cite{Farris-1968} has investigated the relative viscosity of
bimodal suspensions in the framework of a rigorous theoretical
approach, valid only if the coarse particle size is much larger
than the fine particle one (i.e. when $\lambda$, the coarse to
fine particle ratio, is greater than $10$). When this condition is
fulfilled, one can consider that the coarse particles interact
with the Newtonian fine particle suspension, and thence, the
relative viscosity of the bimodal suspension is equal to the
product of the relative viscosities of each monodisperse
suspension. Of course, Farris approach can be generalized to
polydisperse suspensions or to non-Newtonian suspending fluids
provided that estimates of the monodisperse suspension's overall
properties exist. \cite{Stovall-Buil-Such-1987} generalized Farris
approach by taking into account interactions between the particles
of different size using the random close packing's model
of~\cite{Stovall-de-Larrard-Buil-1986} in order to address
problems where scale separation between particles of different
size is not possible. Nevertheless, it is believed that this work
lacks of rigor because Farris reasoning is used by the authors in
a situation where particle size separation is not possible. Later,
\cite{Phan Thien-Pham-1997} developed a multiphase model for
polydisperse suspensions. The idea consists in starting from the
homogeneous linear suspending material (a Newtonian fluid or a
Hookean solid) and in introducing the particles by infinitesimal
volume fraction in the framework of an iterative process
(differential scheme). The first step actually corresponds to the
Einstein approach to the overall properties of a dilute
suspension. Next steps consist in removing a small volume fraction
of the suspending overall medium and in replacing it by the same
volume of particles. Of course, such a process is rigorous only if
particle size separation is possible at each step. Then, this
model is only valid for the low solid volume fractions and hence
is not applicable to concentrated suspensions. By the way,
\cite{Phan Thien-Pham-1997} found that the values of the
rheological properties estimated by this model are much smaller
than those experimentally measured on polydisperse concentrated
suspensions.

Besides theoretical works, experimental studies have also been
performed. \cite{Shapiro-Probstein-1992},
\cite{Probstein-Sengun-Tseng-1994} and \cite{Chang-Powell-1994}
used different types of viscometers to measure the steady
viscosity of bidisperse suspensions. The same general trends are
observed in these works. For a given value of the solid volume
fraction, the viscosity of a bidisperse suspension is lower than
the viscosity of a monodisperse suspension with same solid volume
and the viscosity is a decreasing function of the maximum packing
fraction of the particle size distribution. Interestingly,
\cite{Chang-Powell-1994} also showed that the dimensionless
viscosity $\eta(\phi)/\eta(0)$ of such suspensions is basically a
function of $\phi/\phi_m$ only, where $\phi$ is the particle
volume fraction and $\phi_m$ is claimed to be the maximum packing
fraction of the particle mixture. \cite{Gondret-Petit-1997} have
measured the finite frequency viscosity of bidisperse suspensions.
They also observed that the particle size distribution influences
the dynamic viscosity of bidisperse suspensions, and that the
lower maximum packing density distributions correspond to the
lower dynamic viscosities.

From these studies, it clearly appears that the viscosity of a
polydisperse concentrated suspension is closely related to the
maximum packing fraction of the suspended particles. Accordingly,
the viscosity of a Newtonian suspension can be controlled by
optimizing the particle size
distribution~[\cite{Servais-Jones-Roberts-2002}].

All these studies focused on the influence of the polydispersity
on the viscosity of Newtonian suspensions while polydisperse
non-Newtonian suspensions have been poorly studied. To our
knowledge, this problem has been studied only by few authors and
their studies provide extremely dispersed results. However, as
explained in detail by \cite{Mahaut-Chateau-Coussot-Ovarlez-2008}
and \cite{Mahaut-Mokkedem-Chateau-Roussel-Ovarlez-2008}, the data
reported in these works do not correspond to homogeneous
suspensions of particles interacting only mechanically with the
suspending fluid, which is the subject we are interested in, and
then are not applicable to a wide range of materials. {
\cite{Geiker-etal-2002} studied experimentally the effect of
coarse particles volume fraction on the rheological properties of
self compacting concretes; they studied suspensions of
polydisperse particles of various shapes but of same grading. They
assumed that the effect of aggregates on fresh concrete
rheological properties can be studied by looking to concrete as a
suspension of coarse particles (the aggregates) in a yield stress
fluid (the mortar): coarse particles interact only by
non-Newtonian hydrodynamic forces. Both the fresh concrete and the
mortar behaviors are described by a Bingham law. While
\cite{Geiker-etal-2002} find that the dimensionless rheological
properties for the different studied particles have very different
shapes when plotted vs. the reduced volume fraction
${\phi}/{\phi_m}$, close examination of the results shows that
divergence actually occurs at different ${\phi}/{\phi_m}<1$ for
particles of different shape, in contradiction with the usual
definition of the maximum packing fraction in rheology. In this
study, ${\phi_m}$ was indeed defined as the random close packing
of the dry granular assembly, which may not be relevant for
comparing the rheological data. Moreover, the concrete was tested
in a large gap coaxial cylinder rheometer and its rheological
properties were estimated from the steady-state flow. It is well
known that shear induced migration of particles is likely to occur
in this situation~[\cite{Ovarlez-Bertrand-Rodts-2006}]. Then, the
tested materials may not be homogeneous when the system is at
steady-state and it is not clear that the experimental data of
\cite{Geiker-etal-2002} can be used to estimate the rheological
properties of a well defined material (i.e. a material homogeneous
at the coarse particles scale with a well defined distribution of
particles within the tested volume). The fact that the yield
stress estimates of \cite{Geiker-etal-2002} are much larger than
estimates reported by other authors or in this work also casts
doubt about the validity of their procedure.

\cite{Ancey-Jorrot-2001} have experimentally studied the influence
of adding noncolloidal and non-Brownian particles within a clay
dispersion on the value of the yield stress of the suspension. The
ratio of the particle size was large enough so that the clay
dispersion can be considered as homogeneous at the coarse
particles scale. The authors have chosen to measure the yield
stress of the suspension by means of a slump test in order to
avoid the classical problems encountered with Couette rheometer
(migration of particles, localization of the shear rate,
anisotropy of the material)
[\cite{Coussot-2005,Ovarlez-Bertrand-Rodts-2006}]. They showed
that for well-graded materials, the monodisperse suspension yield
stress does not depend on the particle characteristics (diameter,
material) and that the relative yield stress diverges when the
solid volume fraction value tends toward that of the maximum
packing density. When the coarse particles are polydisperse, the
value of the maximum packing density depends of the size
distribution of the particles. Similarly to what is observed for
Newtonian suspensions, the yield stress diverges for values of the
solid volume fraction depending on the particle size distribution.
The dimensionless yield stresses $\tau_c(\phi)/\tau_c(0)$ measured
for various mixtures of bidisperse particles then seem to collapse
when plotted vs. the reduced volume fraction ${\phi}/{\phi_m}$.
They also observed that for low reduced solid volume fraction, the
yield stress can be a decreasing function of the solid volume
fraction of the coarse particle. This effect was ascribed by the
authors to a depletion phenomena. In the closeness of coarse
particles, the clay particles are expelled from the suspending
fluid. This effect induces an increase of the clay particle
concentration far from the coarse particles which are embedded in
a shell of viscous material (pure water) where the yield stress is
naught. Then, the coarse particles can not contribute to the
overall yield stress and behave as empty pores. Obviously, the
observed decrease of the suspension yield stress originates from
physicochemical effects. We recall that such effects are beyond
the scope of this paper. }

In this study, we are interested in situations where bidisperse
mixtures of noncolloidal particles are dispersed in a yield stress
fluid. It is recalled that yield stress fluids have a solid
viscoelastic behavior below a yield stress; above this yield
stress, they behave as liquids and their flow behavior is often
well fitted to a Herschel-Bulkley law~[\cite{Larson-1999}]. The
influence of monodisperse particles on the behavior of yield
stress fluid has been addressed experimentally
by~\cite{Mahaut-Chateau-Coussot-Ovarlez-2008} and
\cite{Mahaut-Mokkedem-Chateau-Roussel-Ovarlez-2008}, and
theoretically by~\cite{Chateau-Ovarlez-Luu-2008}. The main result
of these studies is that, when model materials are carefully
designed to correspond to the theoretical case of homogeneous and
isotropic distribution of monodisperse hard spheres interacting
only mechanically through a yield stress fluid, there is a
theoretical relationship linking the dimensionless linear
properties of such suspensions to their nonlinear properties that
is in very good agreement with the experimental data. As regards
their elastoplastic properties, this relationship reads
\begin{equation}
 \label{eq:Chateau-Ovarlez-Luu-2008}
\tau_c(\phi)/\tau_c(0) = \sqrt{(1-\phi)
  G^{\prime}(\phi)/G^{\prime}(0)}\text{, }
\end{equation}
$\tau_c$ being the yield stress,
$G^{\prime}$ the elastic modulus, and $\phi$ the particle volume
fraction.
They also observed that
both the elastic modulus and yield stress are monotonically
increasing functions of the solid volume fraction $\phi$ which
seem to diverge when $\phi$ tends toward 0.57. Furthermore, their
data are well fitted to a Krieger-Dougherty { like} equation
\begin{equation}
 \label{eq:Krieger-Dougherty-monodisperse}
G^\prime(\phi) = G^\prime (0) \times
\left(1-{\phi}/{\phi_m}\right)^{-2.5 \phi_m} \text{ with } \phi_m=0.57
\end{equation}
and to its nonlinear generalization obtained by putting
Eq.~\ref{eq:Krieger-Dougherty-monodisperse} into
Eq.~\ref{eq:Chateau-Ovarlez-Luu-2008}
\begin{equation}
 \label{eq:Krieger-Dougherty-monodisperse-generalise}
\tau_c(\phi) = \tau_c(0) \times
\sqrt{(1-\phi)(1-\phi/\phi_m)^{-2.5\phi_m}} \text{ with } \phi_m=0.57
\end{equation}

In this work, we again restrict to situations where there is a
scale separation between the paste microstructure and the
noncolloidal particles in suspension and we focus on the purely
mechanical contribution of the particles to the paste behavior.
For this purpose, we use the experimental procedures described in
\cite{Mahaut-Chateau-Coussot-Ovarlez-2008}. Accordingly, the
experimental investigations reported in this paper focus on the
behavior of the pastes in their solid regime, i.e., on the
influence of particles on the elastic modulus and yield stress.

In Sec.\ref{sec:mat-met}, we briefly present the materials
employed and the experimental setup. Elastic modulus and yield
stress measurements are shown in Sec.~\ref{sec:results}. Then, in
Sec.~\ref{sec:phim}, we present a packing model and show that it
allows to account for the dependence of the rheological properties
on the particle mixture composition. Finally, in
Sec.\ref{sec:analysis}, we combine the packing model estimates
with the experimental data, and we demonstrate that it is possible
to accurately predict the overall properties of the studied
suspensions.

\section{Materials and methods}
\label{sec:mat-met} In this section, we briefly present the main
features of the suspensions we studied and of the experimental
procedures we used for the elastic modulus and yield stress
measurements. It is recalled that these procedures were designed
to study the purely mechanical contribution of an isotropic
distribution of particles to the yield stress fluid behavior. The
interested reader is referred
to~[\cite{Mahaut-Chateau-Coussot-Ovarlez-2008}] for a more
detailed presentation.

\subsection{Materials}
\begin{figure}
\begin{center}%
\scalebox{1}{\includegraphics{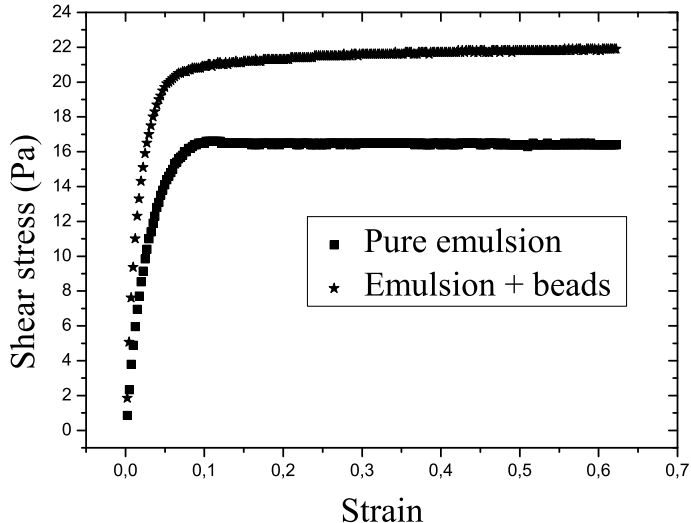}} \caption{Shear
stress vs strain when slowly shearing the material from rest at
$10^{-2}$ $s^{-1}$ for a pure emulsion (squares) and for a
bidisperse suspension of 30\% of 80~$\mu$m polystyrene beads
mixed with 70\% of 315 $\mu$m polystyrene beads with $\phi=0.30$
in the same emulsion (stars).\label{fig:stress_strain}}
\end{center}
\end{figure}

In order to evaluate the purely mechanical influence of the
particles on the behavior of the paste, we designed materials to
ensure scale separation between the matrix (the yield stress
fluid) and the particles. We chose to use only inverse emulsions
as a suspending fluid because
\cite{Mahaut-Chateau-Coussot-Ovarlez-2008} obtained the greater
stability and the more reproducible results with this material in
the monodisperse case. Moreover, the inverse emulsion behavior is
very close to the ideal elastoplastic behavior of a yield stress
fluid~(see~Fig.~\ref{fig:stress_strain}).
The emulsion is a water-in-oil emulsion which microstructure scale is
given by the droplets size, of order 1 $\mu$m from microscope
observations.
{ As the continuous phase, we use a dodecane oil in which a Span 80
emulsifier is dispersed at a 7\% concentration.
The dispersed phase is a 300 g/l CaCl$_2$ solution dispersed in the oil
phase at 6000 rpm for 1 h.}

The particles are spherical beads. We used either polystyrene
beads of density 1.05 g/cm$^3$, or glass beads of density 2.5
g/cm$^3$.
We used bidisperse mixtures of
polystyrene beads of diameter 80 $\mu$m and 315 $\mu$m (particle
size ratio $\lambda = 3.94$) and bidisperse mixtures of glass
beads of diameter 40 $\mu$m and 330 $\mu$m ($\lambda=8.25$). The
bimodal spherical beads are mixed together in the dry state. The
particle mixture composition is defined by the fine particle
proportion $\xi=\Vs/(\Vs+\Vl)$ where $\Vs$ (resp. $\Vl$) denotes
the fine (resp. coarse) particle volume fraction. The particles
are then dispersed in a volume $\Vf$ of the suspending fluid so
that their total volume fraction is $\phi =
(\Vs+\Vl)/(\Vs+\Vl+\Vf)$. Finally, the fluid-particle mixture is
stirred manually in random directions in order to homogenize it
and obtain an isotropic material.
In all cases, the paste yield stress was sufficient to avoid particle
sedimentation in the solid regime.
{ The critical conditions for which a spherical object would fall
  under the action of gravity through a yield stress fluid at rest is
  obtained from the balance between the gravity force, the buoyancy and
  the drag force.
A sphere of radius $a$ and density $\rho$ immersed in a yield stress fluid with yield
stress $\tau_c$ will not move under the action of gravity if $4/3
(\rho - \rho_c) g \pi a^3$ is lower than $4 k_c \pi a^2 \tau_c$
with $k_c \simeq 3.5$, $\rho_c$ the yield stress fluid density and
$g$ the gravity [\cite{Coussot-2005}]. In our experiments, the
critical drag force is at least $30$ times greater than the
gravity force, which ensures stability at rest (thus for the
elastic modulus measurements). When the suspension is sheared at
low shear rate $\dot{\gamma}$, shear-induced sedimentation may
occur with a velocity $V\approx2/9(\rho - \rho_c) g a^2 /
(\tau_c/\dot{\gamma})$ [\cite{Ovarlez-Barral-Coussot-2010}]. Our
yield stress measurements (see below) are performed at very low
shear rate ($\dot{\gamma}=0.01 s^{-1}$): this yields
$V<0.1\mu$m.s$^{-1}$ for all the studied materials, i.e.
shear-induced sedimentation can be considered as negligible over
the duration of the experiments (less than 100 s).}

\subsection{Rheological methods}
The experiments were performed within a vane in cup geometry
(inner diameter $d_i=25$ mm, outer cylinder diameter $d_e=36$ mm,
height $H=45$ mm) on a commercial rheometer (Bohlin C-VOR 2000).
In order to prevent slippage at the walls, we used a six-blade
vane as an inner tool immersed in a outer rough cylinder of
roughness size equivalent to the size of the largest particles. We
measured the elastic modulus $G^{\prime}(\phi)$ through
oscillatory shear experiments at a single frequency in the linear
regime, and the yield stress $\tau_c(\phi)$ through a single
measurement at a small constant velocity ($\dot{\gamma}=0.01
s^{-1}$) on each sample (see an example in
Fig.~\ref{fig:stress_strain}).
{ The (static) yield stress is defined as the shear stress plateau
  in a shear stress vs. shear strain plot (Fig.~\ref{fig:stress_strain}).
It is worth noting that we carefully checked that the same yield
stress is measured whatever the (low) rotational velocity chosen
to drive the inner tool is (i.e. $\dot{\gamma}\leq 0.01 s^{-1}$).}
These methods were shown to provide fair estimates of the
relative evolution of the elastic and yield stress properties with
the particle volume fraction, and to ensure that the studied
suspensions are isotropic and homogeneous
[\cite{Mahaut-Chateau-Coussot-Ovarlez-2008}].

\section{Results}\label{sec:results}
In this section, we present the elastic modulus and yield stress
measurements performed on all the systems we studied, and compare
the results to what was observed in monodisperse suspensions.
Elastic modulus and yield stress measurements of various
monodisperse suspensions of beads embedded in yield stress fluids
were obtained by \cite{Mahaut-Chateau-Coussot-Ovarlez-2008}.
\subsection{Elastic modulus measurement results}
\label{sec:elastic_modulus}

\begin{figure}
\begin{center}%
\scalebox{1}{\includegraphics{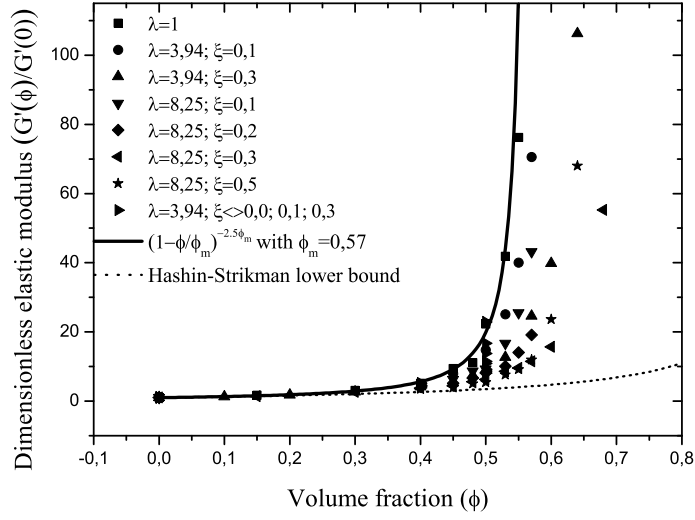}}
\caption{Dimensionless elastic modulus
  $G^{\prime}(\phi,\lambda,\xi)/G^{\prime}(0)$ as a function of the
  particle volume fraction $\phi$ for all the studied materials { (all particle size ratio $\lambda$ and fine particle
proportion $\xi$)}. The solid line is the law
Eq.~\ref{eq:Krieger-Dougherty-monodisperse}. The dotted line is
the Hashin-Shtrikman lower bound Eq.~\ref{eq:hashin_shtrikman
lower bound}. \label{fig:Elas_phi}}
\end{center}
\end{figure}

The results of the elastic modulus measurements performed on all
the suspensions we designed are summarized in Tab.~\ref{tab:sumary
elasticity measurement} and depicted in Fig.~\ref{fig:Elas_phi} as
a function of the solid volume fraction $\phi$.
Two size ratio ($\lambda=3.94$ and $\lambda=8.25$ ) and several fine particle
proportion values $\xi$ were tested.

\begin{table}
\begin{center}
\begin{tabular}{|l|l|l|l|l|l|l|l|}
\hline
\multicolumn{1}{|c|}{$\phi$} & \multicolumn{1}{c|}{$\lambda=1.00$} &
\multicolumn{1}{c|}{$\lambda=3.94$} &
\multicolumn{1}{c|}{$\lambda=3.94$} &
\multicolumn{1}{c|}{$\lambda=8.25$} &
\multicolumn{1}{c|}{$\lambda=8.25$} &
\multicolumn{1}{c|}{$\lambda=8.25$} &
\multicolumn{1}{c|}{$\lambda=8.25$} \\

\multicolumn{1}{|c|}{} & \multicolumn{1}{c|}{} &
\multicolumn{1}{c|}{$\xi=0.1$} & \multicolumn{1}{c|}{$\xi=0.3$} &
\multicolumn{1}{c|}{$\xi=0.1$} & \multicolumn{1}{c|}{$\xi=0.2$} &
\multicolumn{1}{c|}{$\xi=0.3$} & \multicolumn{1}{c|}{$\xi=0.5$} \\

\hline
\multicolumn{1}{|c|}{0.00} & \multicolumn{1}{c|}{1.00} &
\multicolumn{1}{c|}{1.00} & \multicolumn{1}{c|}{1.00} &
\multicolumn{1}{c|}{1.00} & \multicolumn{1}{c|}{1.00} &
\multicolumn{1}{c|}{1.00} & \multicolumn{1}{c|}{1.00} \\
\hline
\multicolumn{1}{|c|}{0.10} & \multicolumn{1}{c|}{} &
\multicolumn{1}{c|}{} & \multicolumn{1}{c|}{1.27} &
\multicolumn{1}{c|}{} & \multicolumn{1}{c|}{} & \multicolumn{1}{c|}{}
& \multicolumn{1}{c|}{} \\
\hline
\multicolumn{1}{|c|}{0.15} & \multicolumn{1}{c|}{1.69} &
\multicolumn{1}{c|}{} & \multicolumn{1}{c|}{1.50} &
\multicolumn{1}{c|}{} & \multicolumn{1}{c|}{} &
\multicolumn{1}{c|}{1.40} & \multicolumn{1}{c|}{} \\
\hline
\multicolumn{1}{|c|}{0.20} & \multicolumn{1}{c|}{} &
\multicolumn{1}{c|}{} & \multicolumn{1}{c|}{1.79} &
\multicolumn{1}{c|}{} & \multicolumn{1}{c|}{} & \multicolumn{1}{c|}{}
& \multicolumn{1}{c|}{} \\
\hline
\multicolumn{1}{|c|}{0.30} & \multicolumn{1}{c|}{3.12} &
\multicolumn{1}{c|}{} & \multicolumn{1}{c|}{2.65} &
\multicolumn{1}{c|}{} & \multicolumn{1}{c|}{} &
\multicolumn{1}{c|}{2.66} & \multicolumn{1}{c|}{} \\
\hline
\multicolumn{1}{|c|}{0.40} & \multicolumn{1}{c|}{5.19} &
\multicolumn{1}{c|}{4.68} & \multicolumn{1}{c|}{4.13} &
\multicolumn{1}{c|}{} & \multicolumn{1}{c|}{} &
\multicolumn{1}{c|}{3.78} & \multicolumn{1}{c|}{} \\
\hline
\multicolumn{1}{|c|}{0.45} & \multicolumn{1}{c|}{9.43} &
\multicolumn{1}{c|}{7.94} & \multicolumn{1}{c|}{5.32} &
\multicolumn{1}{c|}{6.28} & \multicolumn{1}{c|}{5.13} &
\multicolumn{1}{c|}{4.45} & \multicolumn{1}{c|}{3.85} \\
\hline
\multicolumn{1}{|c|}{0.48} & \multicolumn{1}{c|}{11.12} &
\multicolumn{1}{c|}{11.19} & \multicolumn{1}{c|}{7.43} &
\multicolumn{1}{c|}{8.66} & \multicolumn{1}{c|}{6.70} &
\multicolumn{1}{c|}{5.60} & \multicolumn{1}{c|}{5.02} \\
\hline
\multicolumn{1}{|c|}{0.50} & \multicolumn{1}{c|}{22.28} &
\multicolumn{1}{c|}{14.80} & \multicolumn{1}{c|}{9.01} &
\multicolumn{1}{c|}{9.30} & \multicolumn{1}{c|}{8.14} &
\multicolumn{1}{c|}{6.24} & \multicolumn{1}{c|}{5.36} \\
\hline
\multicolumn{1}{|c|}{0.53} & \multicolumn{1}{c|}{41.91} &
\multicolumn{1}{c|}{25.11} & \multicolumn{1}{c|}{12.64} &
\multicolumn{1}{c|}{16.71} & \multicolumn{1}{c|}{10.11} &
\multicolumn{1}{c|}{8.62} & \multicolumn{1}{c|}{7.58} \\
\hline
\multicolumn{1}{|c|}{0.55} & \multicolumn{1}{c|}{76.20} &
\multicolumn{1}{c|}{40.04} & \multicolumn{1}{c|}{15.69} &
\multicolumn{1}{c|}{25.56} & \multicolumn{1}{c|}{14.08} &
\multicolumn{1}{c|}{9.60} & \multicolumn{1}{c|}{9.10} \\
\hline
\multicolumn{1}{|c|}{0.57} & \multicolumn{1}{c|}{} &
\multicolumn{1}{c|}{70.54} & \multicolumn{1}{c|}{24.55} &
\multicolumn{1}{c|}{43.19} & \multicolumn{1}{c|}{19.10} &
\multicolumn{1}{c|}{11.46} & \multicolumn{1}{c|}{11.96} \\
\hline
\multicolumn{1}{|c|}{0.60} & \multicolumn{1}{c|}{} &
\multicolumn{1}{c|}{} & \multicolumn{1}{c|}{39.80} &
\multicolumn{1}{c|}{80.07} & \multicolumn{1}{c|}{31.71} &
\multicolumn{1}{c|}{15.67} & \multicolumn{1}{c|}{23.65} \\
\hline
\multicolumn{1}{|c|}{0.64} & \multicolumn{1}{c|}{} &
\multicolumn{1}{c|}{} & \multicolumn{1}{c|}{106.25} &
\multicolumn{1}{c|}{} & \multicolumn{1}{c|}{60.17} &
\multicolumn{1}{c|}{31.27} & \multicolumn{1}{c|}{68.00} \\
\hline
\multicolumn{1}{|c|}{0.68} & \multicolumn{1}{c|}{} &
\multicolumn{1}{c|}{} & \multicolumn{1}{c|}{} & \multicolumn{1}{c|}{}
& \multicolumn{1}{c|}{} & \multicolumn{1}{c|}{55.29} &
\multicolumn{1}{c|}{} \\
\hline
\end{tabular}
\end{center}
\caption{Dimensionless elastic modulus
  $G^{\prime}(\phi,\lambda,\xi)/G^{\prime}(0)$ as a function of the
  particle volume fraction $\phi$ for all the studied materials { (all particle size ratio $\lambda$ and fine particle
proportion $\xi$)}. \label{tab:sumary elasticity measurement}}
\end{table}

As classically observed for polydisperse viscous suspensions of
different compositions, the experimental points do not fall onto a
single curve.
Nevertheless, we observe that the dimensionless
elastic modulus is an increasing function of the solid volume
fraction $\phi$ when $\lambda$ and $\xi$ are given.
Moreover, all the data points for bidisperse suspensions fall below
the law Eq.~\ref{eq:Krieger-Dougherty-monodisperse}
valid for monodisperse suspensions.
These results are perfectly
consistent with the findings of the literature described in the
introduction.
It is also worth noting that all the experimental
elastic modulus points fall above the Hashin-Shtrikman lower bound
[\cite{Hashin-Shtrikman-1963}]:
\begin{equation}
\label{eq:hashin_shtrikman lower bound}
G^{\prime}(\phi)/G^{\prime}(0)>(2+3\phi)/(2-2\phi)
\end{equation}
which is a theoretical lower bound computed in the general case of
a biphasic material (an infinitely rigid phase embedded in a
linear elastic phase) isotropic both at the microscopic and the
macroscopic scales.

\subsection{Yield stress measurement
results}\label{sec:yield_stress}

We now present the results of the yield stress measurements
performed on the different suspensions we studied. The
experimental data are gathered in Tab.~\ref{tab:sumary yield
stress measurement} and graphically represented in
Fig.~3 as a function of the solid volume
fraction $\phi$.

\begin{table}
\begin{center}
\begin{tabular}{|l|l|l|l|l|l|l|l|}
\hline
\multicolumn{1}{|c|}{$\phi$} & \multicolumn{1}{c|}{$\lambda=1.00$} &
\multicolumn{1}{c|}{$\lambda=3.94$} &
\multicolumn{1}{c|}{$\lambda=3.94$} &
\multicolumn{1}{c|}{$\lambda=8.25$} &
\multicolumn{1}{c|}{$\lambda=8.25$} &
\multicolumn{1}{c|}{$\lambda=8.25$} &
\multicolumn{1}{c|}{$\lambda=8.25$} \\
\multicolumn{1}{|c|}{} & \multicolumn{1}{c|}{} &
\multicolumn{1}{c|}{$\xi=0.1$} & \multicolumn{1}{c|}{$\xi=0.3$} &
\multicolumn{1}{c|}{$\xi=0.1$} & \multicolumn{1}{c|}{$\xi=0.2$} &
\multicolumn{1}{c|}{$\xi=0.3$} & \multicolumn{1}{c|}{$\xi=0.5$} \\
\hline
\multicolumn{1}{|c|}{0.00} & \multicolumn{1}{c|}{1.00} &
\multicolumn{1}{c|}{1.00} & \multicolumn{1}{c|}{1.00} &
\multicolumn{1}{c|}{1.00} & \multicolumn{1}{c|}{1.00} &
\multicolumn{1}{c|}{1.00} & \multicolumn{1}{c|}{1.00} \\
\hline
\multicolumn{1}{|c|}{0.10} & \multicolumn{1}{c|}{} &
\multicolumn{1}{c|}{} & \multicolumn{1}{c|}{1.10} &
\multicolumn{1}{c|}{} & \multicolumn{1}{c|}{} & \multicolumn{1}{c|}{}
& \multicolumn{1}{c|}{} \\
\hline
\multicolumn{1}{|c|}{0.15} & \multicolumn{1}{c|}{1.11} &
\multicolumn{1}{c|}{} & \multicolumn{1}{c|}{} & \multicolumn{1}{c|}{}
& \multicolumn{1}{c|}{} & \multicolumn{1}{c|}{1.11} &
\multicolumn{1}{c|}{} \\
\hline
\multicolumn{1}{|c|}{0.20} & \multicolumn{1}{c|}{} &
\multicolumn{1}{c|}{} & \multicolumn{1}{c|}{1.13} &
\multicolumn{1}{c|}{} & \multicolumn{1}{c|}{} & \multicolumn{1}{c|}{}
& \multicolumn{1}{c|}{} \\
\hline
\multicolumn{1}{|c|}{0.30} & \multicolumn{1}{c|}{1.26} &
\multicolumn{1}{c|}{} & \multicolumn{1}{c|}{1.32} &
\multicolumn{1}{c|}{} & \multicolumn{1}{c|}{} &
\multicolumn{1}{c|}{1.31} & \multicolumn{1}{c|}{} \\
\hline
\multicolumn{1}{|c|}{0.40} & \multicolumn{1}{c|}{1.65} &
\multicolumn{1}{c|}{} & \multicolumn{1}{c|}{1.54} &
\multicolumn{1}{c|}{} & \multicolumn{1}{c|}{} &
\multicolumn{1}{c|}{1.60} & \multicolumn{1}{c|}{} \\
\hline
\multicolumn{1}{|c|}{0.45} & \multicolumn{1}{c|}{2.26} &
\multicolumn{1}{c|}{1.68} & \multicolumn{1}{c|}{1.73} &
\multicolumn{1}{c|}{1.75} & \multicolumn{1}{c|}{1.73} &
\multicolumn{1}{c|}{1.69} & \multicolumn{1}{c|}{1.69} \\
\hline
\multicolumn{1}{|c|}{0.48} & \multicolumn{1}{c|}{2.87} &
\multicolumn{1}{c|}{} & \multicolumn{1}{c|}{2.16} &
\multicolumn{1}{c|}{1.94} & \multicolumn{1}{c|}{1.65} &
\multicolumn{1}{c|}{1.99} & \multicolumn{1}{c|}{1.74} \\
\hline
\multicolumn{1}{|c|}{0.50} & \multicolumn{1}{c|}{3.53} &
\multicolumn{1}{c|}{2.45} & \multicolumn{1}{c|}{2.35} &
\multicolumn{1}{c|}{2.32} & \multicolumn{1}{c|}{2.03} &
\multicolumn{1}{c|}{2.01} & \multicolumn{1}{c|}{1.94} \\
\hline
\multicolumn{1}{|c|}{0.53} & \multicolumn{1}{c|}{5.32} &
\multicolumn{1}{c|}{2.81} & \multicolumn{1}{c|}{2.91} &
\multicolumn{1}{c|}{3.26} & \multicolumn{1}{c|}{2.16} &
\multicolumn{1}{c|}{2.24} & \multicolumn{1}{c|}{2.10} \\
\hline
\multicolumn{1}{|c|}{0.55} & \multicolumn{1}{c|}{8.00} &
\multicolumn{1}{c|}{4.27} & \multicolumn{1}{c|}{} &
\multicolumn{1}{c|}{4.23} & \multicolumn{1}{c|}{2.43} &
\multicolumn{1}{c|}{2.41} & \multicolumn{1}{c|}{2.54} \\
\hline
\multicolumn{1}{|c|}{0.57} & \multicolumn{1}{c|}{} &
\multicolumn{1}{c|}{6.13} & \multicolumn{1}{c|}{4.06} &
\multicolumn{1}{c|}{5.34} & \multicolumn{1}{c|}{2.77} &
\multicolumn{1}{c|}{2.81} & \multicolumn{1}{c|}{2.59} \\
\hline
\multicolumn{1}{|c|}{0.60} & \multicolumn{1}{c|}{} &
\multicolumn{1}{c|}{} & \multicolumn{1}{c|}{4.58} &
\multicolumn{1}{c|}{5.96} & \multicolumn{1}{c|}{} &
\multicolumn{1}{c|}{3.26} & \multicolumn{1}{c|}{5.07} \\
\hline
\multicolumn{1}{|c|}{0.64} & \multicolumn{1}{c|}{} &
\multicolumn{1}{c|}{} & \multicolumn{1}{c|}{7.30} &
\multicolumn{1}{c|}{11.72} & \multicolumn{1}{c|}{7.21} &
\multicolumn{1}{c|}{} & \multicolumn{1}{c|}{9.06} \\
\hline
\multicolumn{1}{|c|}{0.68} & \multicolumn{1}{c|}{} &
\multicolumn{1}{c|}{} & \multicolumn{1}{c|}{} & \multicolumn{1}{c|}{}
& \multicolumn{1}{c|}{} & \multicolumn{1}{c|}{5.64} &
\multicolumn{1}{c|}{} \\
\hline
\end{tabular}
\end{center}
\caption{Dimensionless yield stress
  $\tau_{c}(\phi,\lambda,\xi)/\tau_{c}(0)$ as a function of the
  particle volume fraction $\phi$ for all the studied materials { (all particle size ratio $\lambda$ and fine particle
proportion $\xi$)}. \label{tab:sumary yield stress measurement}}
\end{table}
\begin{figure}
\begin{center}%
\scalebox{1}{\includegraphics{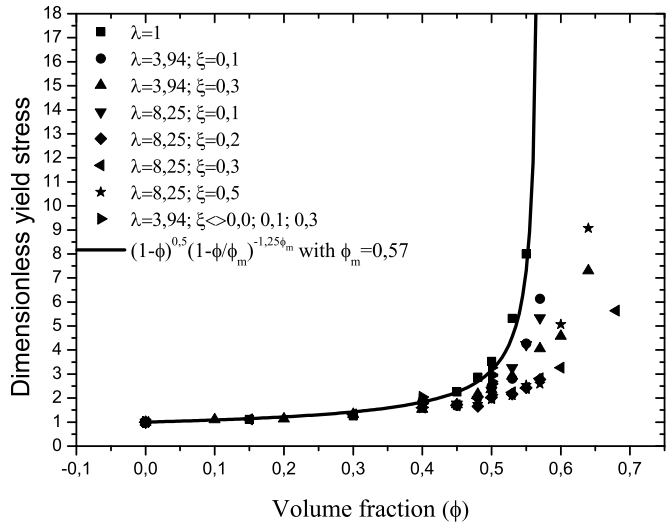}}
\caption{Dimensionless yield stress
  $\tau_{c}(\phi,\lambda,\xi)/\tau_{c}(0)$ as a function of the
  particle volume fraction $\phi$ for all the studied materials { (all particle size ratio $\lambda$ and fine particle
proportion $\xi$)}. The solid line represents the law
$\tau_c(\phi) = \tau_c(0) \sqrt{(1-\phi)(1-\phi/\phi_m)^{-2.5
\phi_m}}$ with $\phi_m=0.57$ proposed
by~\cite{Chateau-Ovarlez-Luu-2008} for the monodisperse
suspensions. \label{fig:seuil_phi}}
\end{center}
\end{figure}

The dimensionless yield stress exhibits the same trends as the
dimensionless elastic modulus.
It is an increasing function of the solid volume fraction when the particle
size ratio $\lambda$ and the fine particle proportion $\xi$ of the
mixture are given.
Moreover, similarly to what was observed for monodisperse suspensions
[\cite{Mahaut-Chateau-Coussot-Ovarlez-2008}], for a given particle
mixture, the yield stress increase with the particle volume
fraction is much lower than the elastic modulus increase.

\section{Divergence of the rheological properties:
The rigidity thres\-hold of contact network}\label{sec:phim}

Although the data shown in Sec.~\ref{sec:results} are scattered,
we observe that all the $G^{\prime}$ and $\tau_c$ curves have
basically the same shape and tend to diverge for a value $\phi_m$
of the volume fraction that seems to depend both on the size ratio
$\lambda$ and on the fine particle proportion $\xi$.
As a first
step towards the modelling of the rheological properties of the
suspensions, it thus seems important to be able to predict the
value of $\phi_m$.
This is the purpose of this section.

\subsection{Packing model}\label{sec:model}

\cite{Mahaut-Chateau-Coussot-Ovarlez-2008} have shown that the
elastic modulus $G^{\prime}$ of an isotropic suspension of
monodisperse spherical beads is well fitted to the
Krieger-Dougherty { like} law
Eq.~\ref{eq:Krieger-Dougherty-monodisperse} with $\phi_m = 0.57$.
{ The quantity $\phi_m$ for which $G^{\prime}$ diverges is most
probably no more than the contact rigidity threshold of the
suspension: if $\phi < \phi_m$ the particles interact only
(hydrodynamically) through the suspending fluid. If $\phi \geq
\phi_m$, the particles are close enough so that there is contact
network spanning the whole material and that interparticle
interactions (contact, friction, elasticity of the beads material
...) play a major role, leading to the apparent divergence of
$G^{\prime}$. } Interestingly, the overall yield stress of
monodisperse suspensions also strongly increases when $\phi$ tends
towards the same value $\phi_m = 0.57$.

At this stage, $\phi_m$ is just a fitting parameter. However,
according to the observed trends of the overall properties of the
bidisperse suspensions in
Sec.~\ref{sec:results}\ref{sec:elastic_modulus} and
Sec.~\ref{sec:results}\ref{sec:yield_stress}, there is strong
evidence that this contact rigidity threshold is closely linked to the
random close packing
[\cite{Chang-Powell-1994,Gondret-Petit-1997,Probstein-Sengun-Tseng-1994,
Frankel-Acrivos-1967}]. This suggests to use packing models to
predict how the value of the contact rigidity threshold vary with the
beads mixture composition. Furthermore, it is largely accepted
that the packing density of an assembly of monodisperse beads
depends upon the small gap interparticle forces and the way the
packing is produced. For example, \cite{Dong-Yang-Zou-Yu-2006}
obtained packing fractions ranging from $0$ to $0.64$ depending on
the conditions between particles ($0.64$ for random close packing
of frictionless cohesionless beads and $0$ for random loose
packing of beads involving strong van der Waals forces). When the
cohesive forces between particles do not dominate over other
forces, the random loose packing fraction, or more precisely, the
sphere packing at its contact rigidity threshold ranges from
$0.5$ to $0.58$~[\cite{Cumberland-Crawford-1987,
Onoda-Liginer-1990,Dong-Yang-Zou-Yu-2006}].
Then, the contact rigidity
threshold $\phi_m = 0.57$ determined experimentally for
monodisperse suspension, assuming it is representative of our
mixture process, may be used as a fixed parameter of a packing
model.

Various models have been proposed in the literature to address
granular mixture problems. For a bidisperse beads mix, the packing
density is a function of the particle size ratio $\lambda$, the
fine particle proportion $\xi$ and the way the packing was
obtained~[\cite{Ben Aïm-Le Goff-1967},
\cite{Dodds-1980,de-Larrard-1999}]. In most of the models of the
literature, the packing process was taken into account through a
scalar index. { In this work, we do not need to model the
mixing process because we always used the same procedure to
prepare the materials. We will first compute the value of the
contact rigidity threshold of our bidisperse packings taking the
monodisperse contact rigidity threshold $\phi_m$ as a free
parameter; the value of $\phi_m$ will then be fixed in the sequel
as the experimentally observed value $0.57$ for our mixing
process.} We chose to use the model of \cite{de-Larrard-1999}
which aims at taking the geometrical interactions between
particles of different size into account. In this model, two
different configurations of bidisperse mixtures are distinguished:
dominant fine particle configuration when coarse particles are
embedded in a fine particle matrix and dominant coarse particle
configuration when fine particles fill the empty spaces between
coarse particles. Furthermore, two geometrical interactions are
taken into account:
\begin{description}
\item[Loosening effect]: When one fine particle is inserted into a
packing of coarse particles (which are thus dominant class) and if
the fine particle is not small enough to locate in the empty space
between the coarse particles,
 there is a loosening of the coarse particle packing which induces a
decrease of the overall coarse particle density.

\item[Wall effect]: When one large particle is inserted into a
local packing of fine particles (dominant fine particle
configuration), the fine particle density is decreased near the
coarse particle. This wall effect induces a decrease of the
overall fine particle density.
\end{description}

\cite{Bournonville-Coussot-Chateau-2004} used a slightly modified
version of the model of~\cite{de-Larrard-1999} to estimate the
packing density of dry polydisperse beads mixtures. They showed
that it is possible to directly use the experimentally measured
packing density associated with the particular packing process
used as a free parameter of the model. Moreover, they
experimentally determined the function describing the loosening
and wall effects described above. While they measured  random
close packing for  monodisperse beads ranging  from $0.59$ to $0.62
$ for particles with diameters between $50 \mu$m and $320 \mu$m,
we used exactly the same equations to estimate the divergence
threshold of bidisperse suspensions, the only difference being the
monodisperse threshold value.
Since we are interested only in bidisperse mixtures, we only give
the relationships allowing to estimate the contact rigidity threshold
for such a suspension. The interested reader is referred to
[\cite{de-Larrard-1999,Bournonville-Coussot-Chateau-2004}] for a
more detailed review of the model.

To compute the contact rigidity threshold of a bidisperse
mixture of particles of respective diameters $\dl$ and $\ds$ with
$\dl>\ds$, one may first calculate the fine dominant particle
density $\phis$ and the coarse dominant particle density $\phil$
defined by
\begin{eqnarray}
\label{eq:phis}
 \phis & = & \frac{\betas}{1-(1-\xi)
 \left(1-\betas+\bsl(\betas-\frac{\betas}{\betal} )\right )}
 \\
\nonumber \\
\label{eq:phil}
 \phil & = & \frac{\betal}{1-\xi \left( 1-\als\betal/\betas
 \right )}
\end{eqnarray}
where $\betas$ (resp. $\betal$) denote the contact rigidity threshold
of the monodisperse fine (resp. coarse) suspension, $\als$ (resp.
$\bsl$) a function describing the loosening (resp. wall) effect
and $\xi$ is the fine particle proportion defined above. In the
sequel, we adopt $\betal = \betas = 0.57$ as determined
experimentally by \cite{Mahaut-Chateau-Coussot-Ovarlez-2008}. The
functions $\als$ and $\bsl$ were determined experimentally by
\cite{Bournonville-Coussot-Chateau-2004}. They only depend upon
the particle size ratio $\lambda= \dl/d_s$.
\begin{eqnarray}
\label{eq:als} \als & = & \left(1-\left(
1-\frac{1}{\lambda}\right)^{1.13}
\right)^{0.57} \\
\nonumber \\
\label{eq:bsl} \bsl & = &\left(1-\left(
1-\frac{1}{\lambda}\right)^{1.79} \right)^{0.82}
\end{eqnarray}
Finally, the bidisperse suspension contact rigidity threshold
is defined by
\begin{equation}
\label{eq:phim} \phi_m=\min(\phil,\phis)
\end{equation}

\begin{figure}
\begin{center}%
\scalebox{1}{\includegraphics{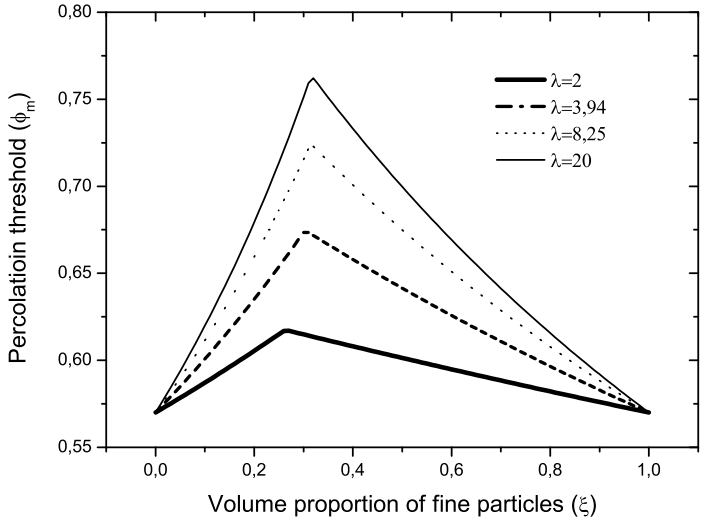}}
\caption{Contact Rigidity threshold $\phi_m$ of bidisperse
suspensions as a function of the fine particle proportion $\xi$,
for several size ratio $\lambda=2$, $\lambda=3.94$, $\lambda=8.25$ and
$\lambda=20$.
\label{fig:bidispersesec}}
\end{center}
\end{figure}

Model predictions are shown in Fig.~\ref{fig:bidispersesec}, where
the contact rigidity threshold $\phi_m$ is plotted as a function of the
fine particle proportion for several values of the size ratio
$\lambda$.

{ Note that whatever the value of the size ratio, the maximum
of the contact rigidity
  threshold  is a singular point of the contact rigidity threshold versus
  fine  particle proportion curve.
  Even if such a singular point was not experimentally observed for
  bidisperse packing of dry particles, it seems to be the hallmark of
  theoretical model distinguishing two regimes
  [\cite{Gondret-Petit-1997,de-Larrard-1999}].
  \cite{de-Larrard-1999} corrected this undesirable
  effect by introducing a scalar index accounting for the packing
  process.
  Other models have been built to predict smooth variations of
  the packing density with respect to the fine particle proportion
  [\cite{Gondret-Petit-1997}].
  To our knowledge, it is not clear that smooth models are
  more accurate than non smooth ones.
  Furthermore, it will be shown in the sequel that
  Eqs.~\ref{eq:phis} to \ref{eq:phim} are accurate enough to
  satisfactorily predict the influence of polydispersity onto both the
  elastic modulus and the yield stress of our suspensions.
  Then, the existence of this singular point seems to be a detail at this stage,
  and does not pose any problem
  in the framework of this study.
}
\subsection{Rheological properties vs Reduced solid
volume fraction}

\begin{figure}
\begin{center}%
\scalebox{1}{\includegraphics{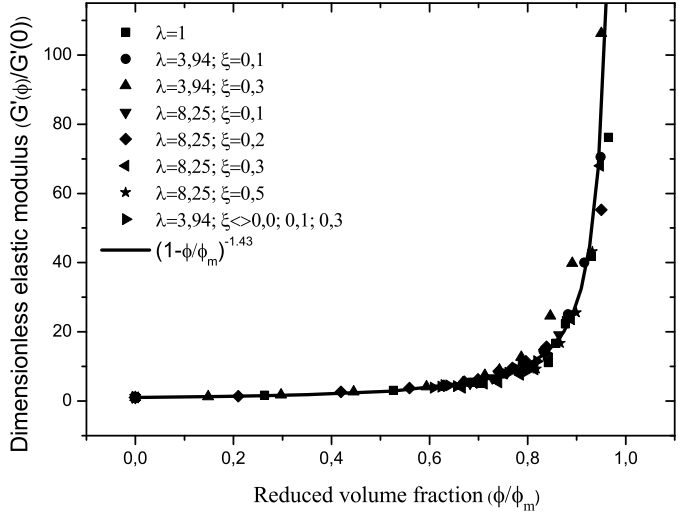}}
\caption{Dimensionless elastic modulus
$G^{\prime}(\phi, \lambda, \xi )/G^{\prime}(0)$
 as a function of the reduced solid volume fraction $\phi/\phi_m$ for
 all bidisperse suspensions { (all particle size ratio $\lambda$ and fine particle
proportion $\xi$). The contact rigidity thresholds $\phi_m$ were
calculated using the modified de Larrard model Eqs.~\ref{eq:phis}
to \ref{eq:phim}.} \label{fig:Elas_phi_normalise}}
\end{center}
\end{figure}

\begin{figure}
\begin{center}%
\scalebox{1}{\includegraphics{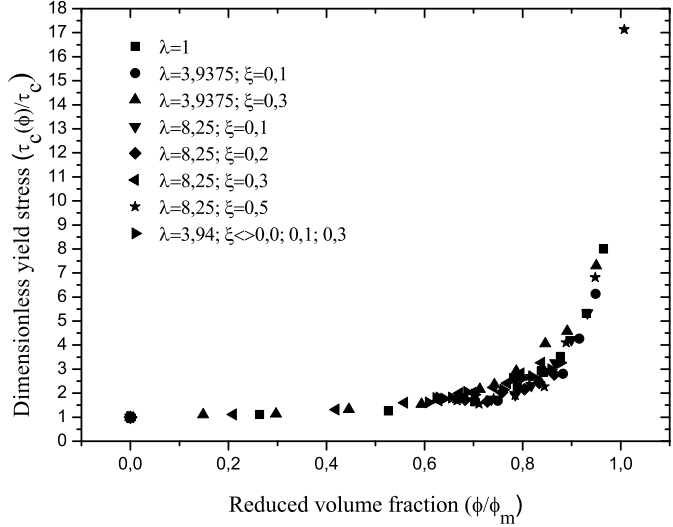}}
\caption{Dimensionless yield stress $\tau_c(\phi, \lambda,
  \xi)/\tau_c(0)$ vs the reduced beads volume fraction $\phi/\phi_m$
  for all bidisperse suspensions { (all particle size ratio $\lambda$ and fine particle
proportion $\xi$). The contact rigidity thresholds $\phi_m$ were
calculated using the modified de Larrard model Eqs.~\ref{eq:phis}
to \ref{eq:phim}.} \label{fig:seuil_phi_normalise}}
\end{center}
\end{figure}

In Sec.~\ref{sec:phim}\ref{sec:model}, we have modelled the value
of the contact rigidity threshold $\phi_m$, i.e. the volume fraction
for which the suspension rheological properties should diverge.
We now propose to plot the dimensionless elastic moduli
$G^{\prime}(\phi,\lambda,\xi)/G^{\prime}(0)$ and the dimensionless yield
stress $\tau_c(\phi,\lambda,\xi)/\tau_c(0)$ of all our bidisperse suspensions
as a function of the predicted reduced solid volume fraction
$\phi/\phi_m$.
These data are presented in Fig.~\ref{fig:Elas_phi_normalise} and
\ref{fig:seuil_phi_normalise}.

We first observe that, when plotted versus $\phi/\phi_m$, the
dimensionless elastic modulus now falls onto a single master
curve. { As the experimental data for monodisperse suspensions
also fall onto this curve, this curve should be fitted by a
Krieger Dougherty like equation. This question is addressed in the
sequel.} It is not obvious that such a property holds for the
dimensionless yield stress, even if the experimental data are less
scaterred than when they are plotted as a function of the solid
volume fraction. We observe that both rheological properties
tend to diverge for $\phi/\phi_m=1$. This last observation is a
strong indication that the packing model we use is able to predict
correctly the impact of the particle size distribution on the
value of the volume fraction for which the rheological properties
diverge.

\section{Analysis and discussion}\label{sec:analysis}

In Sec.~\ref{sec:phim}, we have shown that plotting the
dimensionless rheological properties of bidisperse suspensions as
a function of the reduced solid volume fraction $\phi/\phi_m$
allows to obtain curves diverging when the reduced solid volume
fraction tends toward $1$.
 The purpose of the present section is to provide close-form
estimates for the rheological properties of the bidisperse
suspensions.

\subsection{Elastic modulus}

We have observed that dimensionless elastic moduli of all
bidisperse suspensions plotted as a function of the reduced solid
volume fraction $\phi/\phi_m$ fall onto a single master curve.
Then, by slightly modifying
Eq.~\ref{eq:Krieger-Dougherty-monodisperse} so that the exponent
does not depend on $\phi_m$, we obtain a new estimate for the
elastic modulus of bidisperse suspension
\begin{equation}
\label{eq:prediction of elastic modulus}
G^{\prime}(\phi, \lambda, \xi)/G^{\prime}(0)=(1-\phi/\phi_m)^{-1.43}
\end{equation}
which is shown as a solid line in Fig.~\ref{fig:Elas_phi_normalise}.
Satisfactorily, experimental data are very well fitted
to Eq.~\ref{eq:prediction of elastic modulus}.
{ As we have $1.43 = 2.5 \times 0.57$ for the exponent value in
Eq.~\ref{eq:prediction of elastic modulus}, the Krieger-Dougherty {
  like}
Eq.\ref{eq:prediction of elastic modulus} agrees with
Eq.~\ref{eq:Krieger-Dougherty-monodisperse}. } However, even if a
good agreement between the experimental data and the theoretical
equation is obtained, it has to be noted that first order series
expansion of Eq.~\ref{eq:prediction of elastic modulus} with
respect to the variable $\phi$ yields the \cite{Einstein-1906}
relation
\begin{equation}
\label{eq:einstein}
G^{\prime}(\phi\lambda, \xi)/G^{\prime}(0)=1+2.5 \phi + O^2(\phi)
\end{equation}
only for monodisperse suspensions, i.e. when $\phi_m = 0.57$.
Discrepancy between estimates of the viscosity of concentrated
suspension of solid particles and Einstein's equation is
classical~[\cite{Chang-Powell-1994,Frankel-Acrivos-1967}]. Eilers
formula [\cite{Stickel-Powell-2005}] complies with both the
Einstein equation and high concentration limit, but of course,
this equation does not write as a single function of the
normalized solid volume fraction $\phi/\phi_m$ because
Eq.~\ref{eq:einstein} does not.
As our experimental data are well
fitted to the closed form estimate Eq.~\ref{eq:prediction of
elastic modulus}, we leave this problem aside in this work.

In Fig.~7, we plot the experimentally measured
dimensionless elastic moduli $G^{\prime}(\phi\lambda, \xi
)/G^{\prime}(0)$ as a function of the fine grain proportion $\xi$, for
fixed size ratio $\lambda=3.94$, for two solid volume fraction
$\phi=0.4$ and $\phi=0.5$.
\begin{figure}
\begin{center}%
\scalebox{1}{\includegraphics{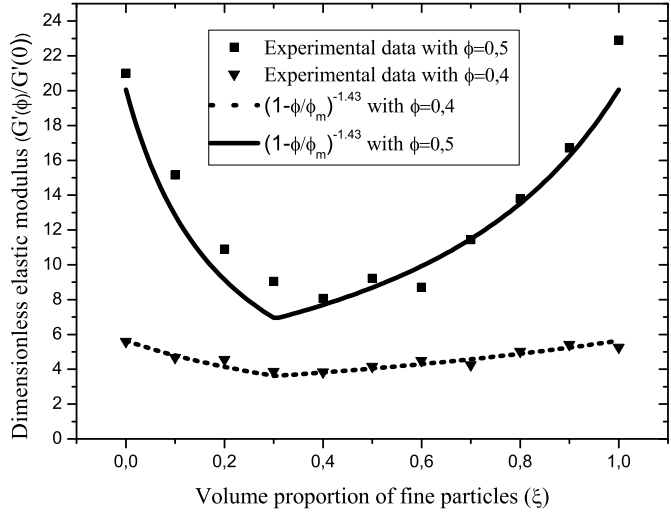}} \caption{Dimensionless
elastic modulus $G^{\prime}(\phi, \lambda, \xi)/G^{\prime}(0)$ vs
fine particle proportion $\xi$ for suspensions of 80 $\mu$m and
315 $\mu$m polystyrene beads with solid volume fraction $\phi=0.4$
(filled triangle) and $\phi=0.5$ (filled square). The dot (resp.
solid) line is Eq.~\ref{eq:prediction of elastic modulus} computed
for $\phi=0.4$ (resp. $\phi=0.5$). The contact rigidity thresholds
$\phi_m$ were calculated using the modified de Larrard model
Eqs.~\ref{eq:phis} to \ref{eq:phim}. \label{fig:elas_xi} }
\end{center}
\end{figure}
These experimental data are  gathered in Tab.~\ref{tab:elas_seuil_xi}.

\begin{table}[h]
\begin{center}
\begin{tabular}{|l|l|l|l|l|l|l|}
\cline{1-3}\cline{5-7}
\multicolumn{1}{|c|}{Elastic Modulus} &
\multicolumn{1}{c|}{$\phi=0.4$} & \multicolumn{1}{c|}{$\phi=0.5$} &
\multicolumn{1}{c|}{} & \multicolumn{1}{c|}{Yield stress} &
\multicolumn{1}{c|}{$\phi=0.4$} & \multicolumn{1}{c|}{$\phi=0.5$} \\
\cline{1-3}\cline{5-7}
\multicolumn{1}{|c|}{$\xi=0.0$} & \multicolumn{1}{c|}{5.60} &
\multicolumn{1}{c|}{21.01} & \multicolumn{1}{c|}{} &
\multicolumn{1}{c|}{$\xi=0.0$} & \multicolumn{1}{c|}{1.91} &
\multicolumn{1}{c|}{3.33} \\
\cline{1-3}\cline{5-7}
\multicolumn{1}{|c|}{$\xi=0.1$} & \multicolumn{1}{c|}{4.68} &
\multicolumn{1}{c|}{15.16} & \multicolumn{1}{c|}{} &
\multicolumn{1}{c|}{$\xi=0.1$} & \multicolumn{1}{c|}{1.84} &
\multicolumn{1}{c|}{2.93} \\
\cline{1-3}\cline{5-7}
\multicolumn{1}{|c|}{$\xi=0.2$} & \multicolumn{1}{c|}{4.56} &
\multicolumn{1}{c|}{10.90} & \multicolumn{1}{c|}{} &
\multicolumn{1}{c|}{$\xi=0.2$} & \multicolumn{1}{c|}{1.81} &
\multicolumn{1}{c|}{2.61} \\
\cline{1-3}\cline{5-7}
\multicolumn{1}{|c|}{$\xi=0.3$} & \multicolumn{1}{c|}{3.87} &
\multicolumn{1}{c|}{9.05} & \multicolumn{1}{c|}{} &
\multicolumn{1}{c|}{$\xi=0.3$} & \multicolumn{1}{c|}{1.94} &
\multicolumn{1}{c|}{2.33} \\
\cline{1-3}\cline{5-7}
\multicolumn{1}{|c|}{$\xi=0.4$} & \multicolumn{1}{c|}{3.84} &
\multicolumn{1}{c|}{8.07} & \multicolumn{1}{c|}{} &
\multicolumn{1}{c|}{$\xi=0.4$} & \multicolumn{1}{c|}{1.61} &
\multicolumn{1}{c|}{2.14} \\
\cline{1-3}\cline{5-7}
\multicolumn{1}{|c|}{$\xi=0.5$} & \multicolumn{1}{c|}{4.16} &
\multicolumn{1}{c|}{9.23} & \multicolumn{1}{c|}{} &
\multicolumn{1}{c|}{$\xi=0.5$} & \multicolumn{1}{c|}{1.79} &
\multicolumn{1}{c|}{2.58} \\
\cline{1-3}\cline{5-7}
\multicolumn{1}{|c|}{$\xi=0.6$} & \multicolumn{1}{c|}{4.51} &
\multicolumn{1}{c|}{8.71} & \multicolumn{1}{c|}{} &
\multicolumn{1}{c|}{$\xi=0.6$} & \multicolumn{1}{c|}{1.77} &
\multicolumn{1}{c|}{2.62} \\
\cline{1-3}\cline{5-7}
\multicolumn{1}{|c|}{$\xi=0.7$} & \multicolumn{1}{c|}{4.25} &
\multicolumn{1}{c|}{11.45} & \multicolumn{1}{c|}{} &
\multicolumn{1}{c|}{$\xi=0.7$} & \multicolumn{1}{c|}{1.81} &
\multicolumn{1}{c|}{2.69} \\
\cline{1-3}\cline{5-7}
\multicolumn{1}{|c|}{$\xi=0.8$} & \multicolumn{1}{c|}{5.03} &
\multicolumn{1}{c|}{13.78} & \multicolumn{1}{c|}{} &
\multicolumn{1}{c|}{$\xi=0.8$} & \multicolumn{1}{c|}{1.80} &
\multicolumn{1}{c|}{2.92} \\
\cline{1-3}\cline{5-7}
\multicolumn{1}{|c|}{$\xi=0.9$} & \multicolumn{1}{c|}{5.43} &
\multicolumn{1}{c|}{16.71} & \multicolumn{1}{c|}{} &
\multicolumn{1}{c|}{$\xi=0.9$} & \multicolumn{1}{c|}{2.04} &
\multicolumn{1}{c|}{2.99} \\
\cline{1-3}\cline{5-7}
\multicolumn{1}{|c|}{$\xi=1.0$} & \multicolumn{1}{c|}{5.26} &
\multicolumn{1}{c|}{22.91} & \multicolumn{1}{c|}{} &
\multicolumn{1}{c|}{$\xi=1.0$} & \multicolumn{1}{c|}{1.94} &
\multicolumn{1}{c|}{3.26} \\
\cline{1-3}\cline{5-7}
\end{tabular}
\end{center}
\caption{Dimensionless elastic modulus $G^{\prime}(\phi, \lambda,
  \xi)/G^{\prime}(0)$ and dimensionless yield $\tau_{c}(\phi, \lambda,
  \xi)/\tau_c(0)$ stress as a function of the fine particle proportion
  $\xi$, for particle size ratio $\lambda=3.94$. \label{tab:elas_seuil_xi}}
\end{table}

Experimental $G^{\prime}(\phi\lambda, \xi)/G^{\prime}(0)$
shows a minimum at a value of $\xi$ close to $0.40$.
This minimum should
thus correspond to the optimal mixture of particles of size ratio
$\lambda=3.94$. The experimental data are also compared with
Eq.~\ref{eq:prediction of elastic modulus} in
Fig.~\ref{fig:elas_xi}, the contact rigidity thresholds being computed
thanks to the model presented in Sec.~\ref{sec:phim}.

The agreement between the experimental data and
Eq.~\ref{eq:prediction
 of elastic modulus} is rather good meaning that the granular packing
model combined with the elastic Krieger-Dougherty equation capture
the essential features of the studied system. However, the
theoretical dimensionless elastic modulus has a minimum at $\xi
=0.30$ while the smallest elastic modulus was measured for
suspension with $\xi$ close to $0.4$. This discrepancy possibly
comes from the packing model which was validated on dry beads
packing experiments, whereas beads are suspended in a fluid in
this work. Finally, close inspection of the data suggest that the
dimensionless elastic modulus vs fine particle proportion curve is
smooth while the tangent to the theoretical curve is discontinuous
at the point where $G^{\prime}(\phi, \lambda, \xi)/G^{\prime}(0)$
reaches its minimal value. As pointed out in Sec.~\ref{sec:phim},
this singularity in the theoretical curve clearly originates from
the packing model. Even if the packing model could be improved to
obtain smooth theoretical curves, it is believed that the model
used in this work is accurate enough given the dispersion of
experimental data.

\subsection{Elastic modulus vs yield stress}

Proposing estimates for the overall properties of a suspension of
particles dispersed in a non-Newtonian fluid is challenging.
However, \cite{Chateau-Ovarlez-Luu-2008} have recently shown that
it is possible to relate the overall yield stress of a suspension
of particles isotropically dispersed in a yield stress fluid to
its overall elastic modulus in its solid regime, provided that the
heterogeneities of the strain rate field over the suspending fluid
domain can be neglected (see
Eq.~\ref{eq:Chateau-Ovarlez-Luu-2008}). To obtain this
relationship, it was assumed that the particles are rigid,
noncolloidal and that there are no physicochemical interactions
between the particles and the paste. It is recalled that our
experimental procedure was designed to fulfill all these
hypotheses [\cite{Mahaut-Chateau-Coussot-Ovarlez-2008}]. { In
this case, the apparent
  viscosity $\eta^{\text{app}}$ of the suspension sheared at a macroscopic shear rate $\dot\gamma$ reads
  \begin{equation}
    \label{eq:etaapp}
    \eta^{\text{app}} (\phi, \dot \gamma) = \frac{G(\phi)}{G(0)} \times
    \eta^{\text{app}}_{0} (\dot \gamma_{\text{eff}})
  \end{equation}
where $\eta^{\text{app}}_0$ is the apparent viscosity of the
suspending fluid sheared at an effective strain rate $\dot
\gamma_{\text{eff}}$ that accounts for the shear rate locally
experienced by the pure suspending fluid, and $G(\phi)$ is the
elastic modulus of the same suspension of particles (i.e. same
size and same position) dispersed in an elastic material.
Eq.~\ref{eq:etaapp} simply says that in a nonlinear
(non-Newtonian) medium, the viscosity increase of the suspension
with the particle volume fraction $\phi$ is the same as in
a linear (Newtonian or elastic) medium whose viscosity would be
the apparent viscosity $\eta^{\text{app}}_{0} (\dot
\gamma_{\text{eff}})$ of the sheared suspending fluid; in this
case, an additional dependence of the apparent viscosity on $\phi$
comes from the $\dot \gamma_{\text{eff}}$ dependence on $\phi$. As
the solid particles do not deform, the shear correction factor $
{\dot \gamma_{\text{eff}}}/{\dot \gamma}$ is greater than one. It
can be shown in the framework of a rigorous upscaling approach to
this problem that the optimal estimate of this quantity reads
[\cite{Chateau-Ovarlez-Luu-2008}]
\begin{equation}
  \label{eq:localization}
 {\dot \gamma_{\text{eff}}}/{\dot \gamma} = \sqrt{
   \frac{G(\phi)/G(0)}{(1-\phi)}}
\end{equation}
in which the quantity $1-\phi$ appears because the shear rate
experienced by the suspending fluid is linked to the overall shear
rate by means of an average equation over the interstitial fluid
only. This explains why nonlinear properties cannot depend simply
on the reduced solid volume fraction $\phi/\phi_m$.

This approach allows in particular predicting the value
$\tau_c(\phi)/\tau_c(0)$ of the dimensionless yield stress of
suspensions of noncolloidal particles in yield stress fluids. The
apparent viscosity of a perfect plastic yield fluid indeed reads
$\tau_c/{\dot\gamma}$. Putting this equation into
Eq.~\ref{eq:etaapp} combined with Eq.~\ref{eq:localization}
finally yields Eq.~\ref{eq:Chateau-Ovarlez-Luu-2008}.

This analysis still allows to physically explain why the relative yield stress
increase is lower than the relative elastic modulus increase.
Indeed, while the relative apparent viscosity of the suspension is
equal to its relative elastic modulus, the shear rate experienced by
the suspending fluid is greater than the overall shear rate prescribed
to the suspension.
As the apparent viscosity of a perfect yield stress fluid is a
decreasing function of the shear rate, the localization of the shear
rate lowers the consolidating effect of adding solid particles.}

\begin{figure}
\begin{center}
\scalebox{1}{\includegraphics{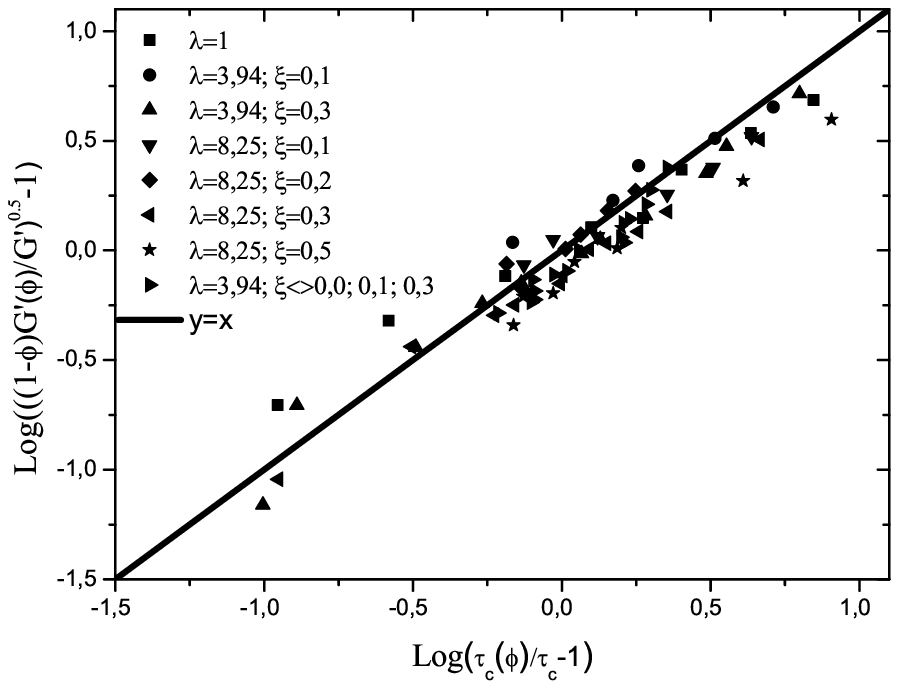}}
\caption{Dimensionless yield stress $\tau_c(\phi, \lambda,
  \xi)/\tau_c(0)$ as a function of $\sqrt{(1-\phi)G^{\prime}(\phi,
    \lambda, \xi)/G^{\prime}(0)}$ for all the bidisperse suspensions
  studied { (all particle size ratio $\lambda$ and fine particle
proportion $\xi$)}.
   The figure's coordinates were chosen so that the $y=x$ line
   represents the theoretical relation Eq.~\ref{eq:Chateau-Ovarlez-Luu-2008}.
\label{fig:relation_elasticity_seuil}}
\end{center}
\end{figure}

We have plotted in Fig.~\ref{fig:relation_elasticity_seuil} the
dimensionless yield stress $\tau_c(\phi, \lambda, \xi)/\tau_c(0)-1$ as a
function of the dimensionless quantity
$\sqrt{(1-\phi)G^{\prime}(\phi, \lambda, \xi)/G^{\prime}(0)}-1$ in logarithmic
coordinates for all the systems we studied in order to check that
Eq.~\ref{eq:Chateau-Ovarlez-Luu-2008} is still valid.
We observe an excellent agreement between our experimental results
and the micromechanical estimate
Eq.~\ref{eq:Chateau-Ovarlez-Luu-2008} (which is plotted as a
straight line $y=x$ in these coordinates).

These results show that
the experimental data are consistent with the hypothesis that a
uniform estimate of the strain rate over the suspending fluid
domain allows to accurately estimate the overall yield stress of a
bidisperse suspension from its overall elasticity.

\subsection{Yield stress}

\begin{figure}
\begin{center}
\scalebox{1}{\includegraphics{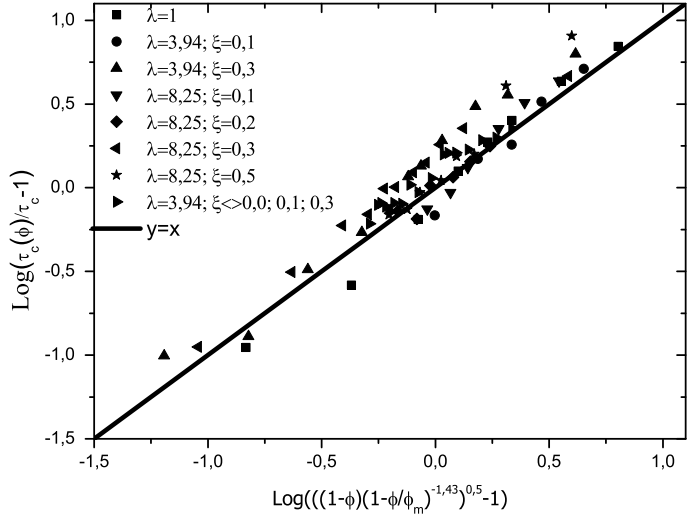}}
\caption{Dimensionless yield stress $\tau_c(\phi, \lambda,
  \xi)/\tau_c(0)$ as a function of
  $\sqrt{(1-\phi)(1-\phi/\phi_m)^{-1.43}}$ for all the bidisperse
  suspensions studied { (all particle size ratio $\lambda$ and fine particle
proportion $\xi$)}. The figure's coordinates were chosen so that
  the $y=x$ line represents the theoretical relation
  Eq.~\ref{eq:prediction of yield stress}. \label{fig:fig9}}
\end{center}
\end{figure}

Putting the estimate Eq.~\ref{eq:prediction of elastic modulus}
for the overall linear properties of the suspension into
Eq.~\ref{eq:Chateau-Ovarlez-Luu-2008} yields the estimate for the
overall yield stress of the suspension:
\begin{equation}
\label{eq:prediction of yield stress}
\tau_c(\phi, \lambda, \xi)/\tau_c=\sqrt{(1-\phi)(1-\phi/\phi_m)^{-1.43}}
\end{equation}
We have plotted in Fig.~\ref{fig:fig9} the
experimental dimensionless yield stress $\tau_c(\phi, \lambda,
\xi)/\tau_c(0)-1$ as a function of the dimensionless quantity
$\sqrt{(1-\phi)(1-\phi/\phi_m)^{-1.43}} -1$ in logarithmic
coordinates for all the systems we studied in order to check that
Eq.~\ref{eq:prediction of yield stress} accurately describes the
experimental data.

Even if the fit is not perfect, it is believed that
Eq.~\ref{eq:prediction of yield stress} is accurate enough to predict
the yield stress of the bidisperse suspensions studied in this paper.

\begin{figure}
\begin{center}
\scalebox{1}{\includegraphics{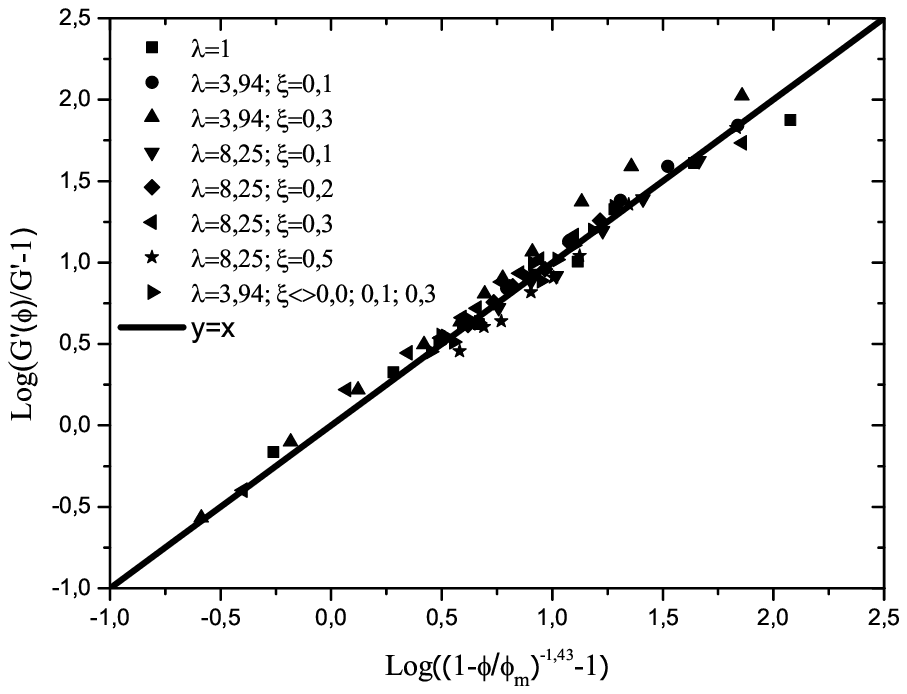}}
\caption{Dimensionless elastic modulus $G^{\prime}(\phi, \lambda,
  \xi)/G^{\prime}(0)$ as a function of $(1-\phi/\phi_m)^{-1.43}$ for
all the bidisperse suspensions studied { (all particle size
ratio $\lambda$ and fine particle proportion $\xi$)}. The figure's
coordinates were chosen so that the $y=x$ line represents the
theoretical relation Eq.~\ref{eq:prediction of elastic modulus}.
\label{fig:fig10}}
\end{center}
\end{figure}

In order to compare the accuracy of Eq.~\ref{eq:prediction of yield
  stress} as an estimate of the overall yield stress of the suspensions with
the accuracy of the elastic modulus estimate Eq.~\ref{eq:prediction of
  elastic modulus}, we have plotted in Fig.~\ref{fig:fig10} the experimental
dimensionless elastic modulus as a function of
$(1-\phi/\phi_m)^{-1.43}$.

At first sight, both estimates seem to accurately fit the
experimental data.
{
To quantitatively assess this point, we have computed the least square
error for both the dimensionless yield stress
(Fig.~\ref{fig:fig9}) and dimensionless elastic modulus
(Fig.~\ref{fig:fig10}).
For quantity $y$ the least square error reads
\begin{equation}
  \label{eq:defErr}
  \text{Err}_y = \frac{1}{N} \sum_{i=1}^{N} \left(y_i^{\text {exp}}
  -y_i^{\text{th}}\right)^2
\end{equation}
with $N$, the number of measured points, $y_i^{\text {exp}}$, the
$i$th experimentally measured point and $y_i^{\text {th}}$ the
associated theoretical value.
We found $\text{Err}_{G} = 0.0071$ for the elastic modulus estimate
and $\text{Err}_{\tau_c} = 0.018$ for the yield stress estimate.
Both computed errors being of the same order of magnitude, it can be
concluded that the yield stress estimate Eq.\ref{eq:prediction of
  yield stress} is quite as precise than the elastic modulus estimate
Eq.~\ref{eq:prediction of elastic modulus}.
}
\section{Conclusion}

We have studied the elastic modulus and yield stress of an
isotropic bidisperse suspension of noncolloidal particles in a
yield stress fluid. We focused on the purely mechanical
contribution of the noncolloidal particles to the overall
properties of the yield stress fluid. To do this, we used
materials and procedures designed
by~\cite{Mahaut-Chateau-Coussot-Ovarlez-2008} to study the case of
monodisperse suspensions. We observed that, as it is classically
observed for particles suspended in a Newtonian fluid, the elastic
modulus and yield stress of bidisperse suspensions are lower than
the same quantities measured for monodisperse suspension of same
solid volume fraction. We showed that the equation $\tau_c(\phi,
\lambda,
\xi)/\tau_c(0)=\sqrt{(1-\phi)G^{\prime}(\phi)/G^{\prime}(0)}$
of~\cite{Chateau-Ovarlez-Luu-2008} linking the overall yield
stress, the overall elastic modulus and the solid volume fraction
still applies when bidisperse suspensions are considered.
Additionally, we showed that the effect of the particle size
heterogeneity onto the overall rheological properties of the yield
stress suspension can be described by means of a packing model
developed to estimate random loose packing of assemblies of dry
particles. We have finally proposed closed form estimates for both
the elastic modulus and the yield stress. This shows that it is
sufficient to determine $\phi_m$ and the dependence of the
elastic modulus of monodisperse suspension on $\phi$ to predict the
behavior of
bidisperse suspensions; this should remain true for more complex
polydisperse cases. An extension of this study to the cases of
anisotropic particle distribution and more complex polydisperse
suspensions is planned for the future.

\end{document}